\documentclass{article}
\usepackage[preprint]{spconf}  % preprint: show page numbers; drop for final submission
\usepackage{amsmath,amssymb,mathtools,graphicx,booktabs}
\usepackage{hyperref}
\usepackage[capitalize,noabbrev]{cleveref}
\crefname{figure}{Fig.}{Figs.}
\Crefname{figure}{Fig.}{Figs.}

\graphicspath{{figures/}{.}}

\newcommand{\TIMECONSTANT}{\mu}
\newcommand{\DEV}{ \sigma}

\title{FRAME BOUNDS AND BANDWIDTH TILING FOR TIME-CAUSAL BANDPASS WAVELETS}

\name{Jens E. Pedersen$^{\star}$, Tony Lindeberg$^{\dagger}$, Peter Gerstoft$^{\star}$\thanks{Support from the Novo Nordisk Foundation (NNF24OC0089302) and the Swedish Research Council (2022-02969) is gratefully acknowledged. Code: \url{https://github.com/jegp/swavelet}.}}
\address{$^{\star}$Department of Electrical and Photonics Engineering, Technical University of Denmark, Denmark\\
    $^{\dagger}$Department of Computational Science and Technology, KTH Royal Institute of Technology, Sweden}

\begin{document}
\ninept
\maketitle

\begin{abstract}
  Wavelets and frames that provide strong guarantees for signal representations are a cornerstone in signal processing theory, but have long been restricted to non-causal settings.
  Recent work extended wavelet analysis to time-causal systems that operate on streaming signals, with exact reconstruction and time-recursive implementations.
  However, the frame guarantees that make such representations reliable under noise and quantization are uncharacterized.
  We derive closed-form frame bounds and conditioning on band-limited domains, together with peak frequencies, constant-Q bandwidth tiling, and spectral decay rates for time-causal bandpass wavelets.
  Our work provides design rules for the scale ratio and channel count, which we validate numerically against discrete implementations of the causal wavelets in terms of recursive filters.
\end{abstract}

\begin{keywords}
Wavelet frames, time-causal scale space, frame bounds, filter banks, constant-Q analysis
\end{keywords}
\vspace{-0.1em}
\section{Introduction} \label{intro}
\vspace{-0.1em}
When analyzing signals in real-time for streaming applications, multi-scale analysis, such as wavelets, requires causal filters that only access the past.
Time-causal wavelets were defined early \cite{szu1992causal}, and \cite{lindeberg2025time} recently recast them as bandpass wavelets grounded in scale-space theory \cite{koenderink1984structure}.
These new wavelets define filterbanks as scaled and shifted versions of time-causal kernels \cite{lindeberg2025time}, amenable to real-time applications, but without exact frame bounds, frame conditioning, or quality factor analyses.

Bandpass systems, defined as differences between non-negative kernels, rest on prior work on Laplacian pyramids \cite{do2003framing} and Calder\'on-sum analysis \cite{daubechies1992ten,mallat2009wavelet}; recent work has extended the causal bandpass kernels from \cite{lindeberg2025time} to sparse spiking neural systems \cite{pedersen2026scale}.
Two bandpass wavelets were introduced in \cite{lindeberg2025time}: the Difference-of-Time-causal-limit-kernels (DoT) and a variant of the previously studied Difference-of-Gaussians (DoG).
We build on these by contributing:
%\begin{itemize}
\\%    \item 
1)    The Difference-of-truncated-Exponentials (DoE) as a simplified case of the DoT wavelet.
\\%       \item 
2)    Frame bounds and conditioning for the DoG, DoT, and DoE wavelets, in closed form for the causal families.
\\%       \item 
3)    Bandwidth and constant-Q tiling analysis for the DoG, DoT, and DoE wavelets.
\\%      \item 
4)    Numerical validation of the closed forms against discrete implementations of the causal wavelets as recursive filters.
%\end{itemize}

\vspace{-0.1em}
\section{Scale spaces, Wavelets, and frames}
\vspace{-0.1em}

%\subsection{Scale spaces} \label{scale_spaces}
Scale-space theory parameterizes a signal $f\colon \mathbb{R} \to \mathbb{R}$ over a scale parameter $\DEV \in \mathbb{R}_+$ by convolving with smoothing kernels \cite{koenderink1984structure, lindeberg1994scalespace}
\vspace{-0.1em}
\begin{equation} \label{eq:scale-space}
    L(t;\ \DEV) = h(t;\ \DEV) * f(t), \qquad L(t;\ 0) = f(t).
    \vspace{-0.1em}
\end{equation}
The scale-space representation $L$ is scale covariant, $L'(s \, t; s \, \DEV) = L(t; \DEV)$ for any temporal scaling factor $s > 0$ \cite{lindeberg2023time}.
Admissible kernels are the variation-diminishing Polya frequency functions \cite{schoenberg1948variationdiminishing}: over non-causal domains the unique one forming a continuous semi-group is the Gaussian $h_{\rm Gauss}(t; \DEV)$ \cite{lindeberg1994scalespace}, while causality admits only the one-sided {\it truncated exponential}
\vspace{-0.1em}
\begin{equation} \label{eq:truncated_exponential}
    h_{\rm exp}(t, \TIMECONSTANT) = \TIMECONSTANT^{-1} \exp{(-t / \TIMECONSTANT)} \ \text{for}\ t > 0,\ \text{else}\ 0,
    \vspace{-0.1em}
\end{equation}
with time constant (and delay) $\TIMECONSTANT$.
A logarithmic spread of scale levels with ratio $c > 1$ (default $c = \{\sqrt{2}, 2\}$),
\vspace{-0.1em}
\begin{equation} \label{eq:scale_level_spread}
    \TIMECONSTANT_{k} = c \, \TIMECONSTANT_{k-1},
    \vspace{-0.1em}
\end{equation}
retains self-similarity, and cascading truncated exponentials with such geometrically spaced time constants builds a time-causal scale space that obeys scale covariance for $s = c^j$, $j \in \mathbb{Z}$ \cite{lindeberg2023time, lindeberg2025time}.

% \vspace{-0.1em}
%\subsection{Wavelets and frames} \label{wavelets_frames}
Wavelets decompose a signal $f(t)$ into scale and shift components, $\langle f(t), \psi(t; a, b) \rangle$ with scale $a$ and temporal shift $b$ \cite[p.~24]{daubechies1992ten}.
For $\psi$ to serve as a mother wavelet, it must satisfy the admissibility criterion \cite[(2.4.1)]{daubechies1992ten}
\vspace{-0.1em}
\begin{equation} \label{eq:wavelet_admissibility}
    C_\psi = \int_0^\infty \frac{|\hat{\psi}(\omega)|^2}{\omega} \, \mathrm{d}\omega < \infty,
    \vspace{-0.1em}
\end{equation}
where $\hat{\psi}$ is the Fourier transform of $\psi$, implying the zero-mean condition $\int_{-\infty}^\infty \psi(t) \, {\rm d}t = 0$ \cite[(1.2.3)]{daubechies1992ten}.
A family of wavelets $\psi_j$, indexed by $j \in J$, is a frame if there exist bounds $0 < A \leqslant B < \infty$ such that for all $f$ in a Hilbert space \cite[(3.1.2)]{daubechies1992ten}
\vspace{-0.1em}
\begin{equation} \label{eq:frame_bounds}
    A \, \|f\|^2 \leqslant \sum_{j \in J} |\langle f, \psi_j\rangle |^2 \leqslant B \, \|f\|^2.
    \vspace{-0.1em}
\end{equation}
The frame is tight when $A = B$ and overcomplete when $A < B$.
The frame operator $\Lambda\colon f \mapsto (\langle f, \psi_j\rangle)_j$ \cite[(5.2)]{mallat2009wavelet} analyzes a signal into frame coefficients.
Its adjoint $\Lambda^*$ synthesizes coefficients back into a signal, and the composition $\Lambda^*\Lambda\colon L_2(\mathbb{R}) \to L_2(\mathbb{R})$ is self-adjoint and positive-definite with eigenvalues in $[A, B]$.

For a filterbank of shift-invariant channels, the bounds are given by the energy capture function \cite[Thm.~5.11]{mallat2009wavelet}
\vspace{-0.1em}
\begin{equation} \label{eq:energy_capture_def}
    S(\omega) = \sum_{j \in J} |\hat{\psi}_j(\omega)|^2,
    \vspace{-0.1em}
\end{equation}
with $A = \inf_\omega S(\omega)$ and $B = \sup_\omega S(\omega)$.
The dual frame then reconstructs any signal $f$ exactly, with the conditioning $B/A$ governing stability \cite[Ch.~3]{daubechies1992ten}.

\section{Scale-covariant bandpass representations}
\begin{figure*}
    \centering
    \includegraphics[width=.95\linewidth]{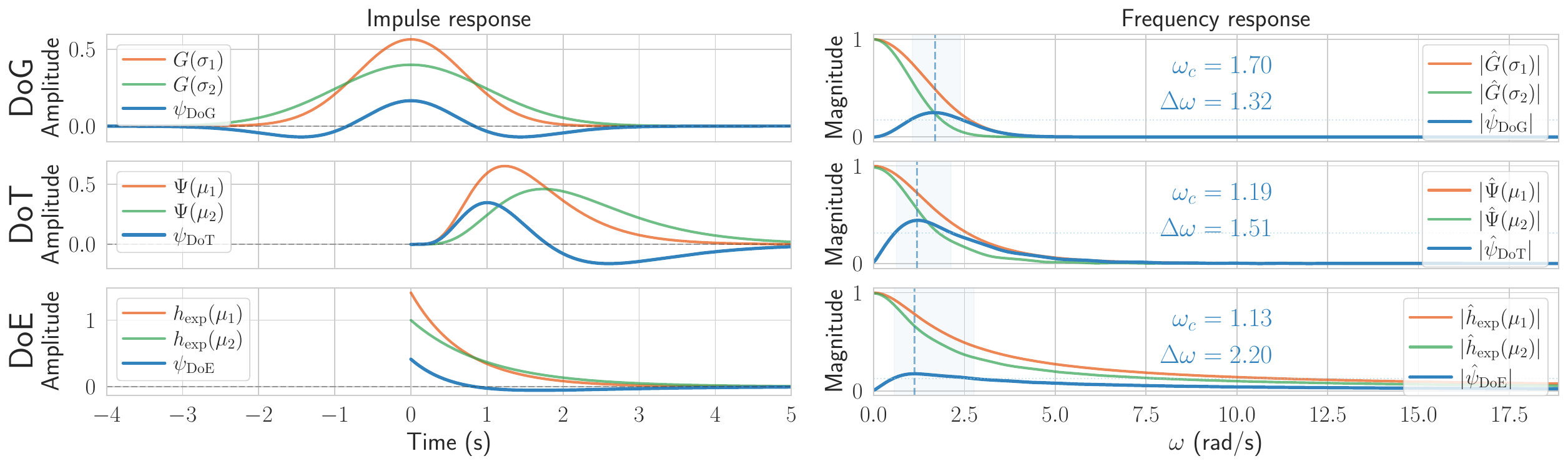}
    \vspace*{-0.4em}
    \caption{Impulse and frequency responses to a single $\delta$ for the DoG \eqref{eq:dog_wavelet}, DoT \eqref{eq:dot_wavelet}, and DoE \eqref{eq:doe_wavelet} wavelets at $c=\sqrt{2}$, for a neighboring scale pair.
    Dashed lines mark each family's peak frequency, cf. \eqref{eq:dog_peak}, \eqref{eq:dot_peak_sandwich} and \eqref{eq:doe_peak}.
    Shading marks its $-3$ dB band $(\omega_-, \omega_+)$ \eqref{eq:constant_quality}.}
    \label{fig:bandpass}
    % \vspace{-1mm}
\end{figure*}
A multi-scale encoder must split into discrete scale channels, be covariant under temporal rescaling, and remove constant components so the channels admit wavelet frames.
Following the Laplacian pyramid \cite{BA83-COM}, scale-space representations \cite{koenderink1984structure}, multi-resolution wavelet frames \cite{mallat1989theory}, and the time-causal bandpass representation in \cite{lindeberg2025time}, we obtain all three from adjacent differences of $L$ \eqref{eq:scale-space} with spacing \eqref{eq:scale_level_spread},
\vspace{-0.1em}
\begin{equation} \label{eq:difference-of-reconstruction}
    \Delta L(t; \DEV_k, c)\! =\! L(t; \DEV_k, c) - L(t; \DEV_{k-1}, c)\! =\! \psi(t;\DEV_k,c) * f(t),
    \vspace{-0.1em}
\end{equation}
which is covariant by construction and vanishes at $\omega = 0$; for the Gaussian it approximates the Laplacian-of-Gaussian \cite{marr1980theory}.
Summing the bandpass channels with a lowpass residual at the coarsest scale $\DEV_K$ reconstructs the signal exactly \cite[(104)]{lindeberg2025time},
\vspace{-0.1em}
\begin{equation} \label{eq:scale_bandpass_reconstruction}
    \tilde{f}(t) = h(t; \DEV_K) * f(t) - \sum^K_{k=1} \psi(t; \DEV_k, c) * f(t),
    \vspace{-0.1em}
\end{equation}
from a frame with $K+1$ channels ($K$ bandpass and one lowpass, $\sigma_0$ as the fastest timescale), whose residual is the scaling function of a finite-level multi-resolution analysis \cite[(9)]{mallat1989theory}.

We use three candidate bandpass kernels $\psi$; the first, based on the Gaussian, is the \textbf{difference-of-Gaussians (DoG)}
\vspace{-0.1em}
\begin{equation} \label{eq:dog_wavelet}
    \psi_{\rm DoG}(t;\DEV_k,c) = h_{\rm Gauss}(t;\DEV_k) - h_{\rm Gauss}(t;\DEV_{k-1}),
    % \vspace{-0.1em}
\end{equation}
which is admissible \eqref{eq:wavelet_admissibility} since $\int_{-\infty}^\infty h_{\rm Gauss}(t;\DEV) \, {\rm d}t = 1$ gives $\int_{-\infty}^\infty \psi_{\rm DoG}(t;\DEV_k, c) \, {\rm d}t = 0$.

For time-causal scale representations, we rely on $h_{\rm exp}(t,\mu)$ \eqref{eq:truncated_exponential} to map the past onto a complete axis \cite{koenderink1988scale} and to provide a realizable computational primitive \cite{pedersen2025covariant}.
However, it lacks the smoothness and scale-covariance of the Gaussian, because its impulse response peaks at $t=0^+$ and decays exponentially.
This family does not guarantee non-creation of new structures from finer to coarser scales, so the simplifying properties to coarser scales are not guaranteed.
The time-causal limit kernel \cite{lindeberg2023time, lindeberg2025time} restores the variation diminishing property by cascading infinitely many $h_{\rm exp}$ stages, recovering scale covariance and the cascade-smoothing property of the Gaussian while remaining causal.
The idealized time-causal limit kernel is constructed by convolving an infinite number ($N = \infty$) of truncated exponential kernels with mean and standard deviation $\mu_n = c^{-n} \sqrt{c^2-1} \, \sigma$ according to \cite[(8, 10)]{lindeberg2023time}
%\vspace{-0.1em}
\begin{equation} \label{eq:time-causal_limit_kernel_convolution}
    \begin{aligned}
        h_\Psi(t; \DEV, c) & = *_{n=1}^{N = \infty}\ h_{\rm exp}(t; \TIMECONSTANT_n), \\[-0.3em]
        \text{mean}[h_\Psi]    & = \sum_{n=1}^{N = \infty} \mu_n = \sigma, \quad
        {\rm std}[h_\Psi] = \sigma = \sqrt{\sum_{n=1}^{\infty} \mu_n^2}.
    \end{aligned}
    \vspace{-0.1em}
\end{equation}
Since $h_\Psi$ is based on the time-causal truncated exponential \eqref{eq:truncated_exponential}, we can insert its scale-space representation
$    L(t; \DEV, c) = h_\Psi(t; \DEV, c) * f(t)$
as a bandpass filter \eqref{eq:difference-of-reconstruction} to construct a wavelet based on the \textbf{difference-of-time-causal-limit-kernels (DoT)} at scale $k$ \cite[(100)]{lindeberg2025time}
\begin{equation} \label{eq:dot_wavelet}
    \psi_{\rm DoT}(t; \DEV_k, c) = h_\Psi(t; \DEV_k, c)  - h_\Psi(t; \DEV_{k - 1}, c).
\end{equation}
If truncating to $N=1$ kernels in \eqref{eq:time-causal_limit_kernel_convolution}, this degenerates to a single $h_{\exp}$ term \eqref{eq:truncated_exponential},
$    L(t; \TIMECONSTANT) = h_{\rm exp}(t; \TIMECONSTANT) * f(t)$,
from which we construct the \textbf{difference-of-truncated-exponentials (DoE)} wavelet:
\begin{equation} \label{eq:doe_wavelet}
    \psi_{\rm DoE}(t; \TIMECONSTANT_k) = h_{\rm exp}(t; \TIMECONSTANT_{k}) - h_{\rm exp}(t; \TIMECONSTANT_{k-1}).
\end{equation}
Signal reconstruction for both the DoE and DoT wavelets follows from \eqref{eq:scale_bandpass_reconstruction}.
Both the DoT and DoE wavelets satisfy the admissibility criterion \eqref{eq:wavelet_admissibility}: each smoothing kernel is a probability distribution integrating to one, so each difference integrates to $1 - 1 = 0$.
\Cref{fig:bandpass} visualizes the impulse and frequency responses for the DoG, DoT, and DoE wavelets.
The DoG is non-causal and can ``react'' backwards in time, while the DoT and DoE will necessarily lag behind.
\vspace{-0.5em}
\section{Spectral characterization} 
\label{spectral}
\vspace{-0.2em}
For each wavelet, we read the frame bounds, peak frequency, $-3$dB bandwidth, and high-frequency decay from a single squared magnitude $|\widehat\psi|^2$.
The bounds follow from the energy capture $S(\omega)$ \eqref{eq:energy_capture_def} as $B = \sup_\omega S$ and $A = \inf_{\omega\in[0,\Omega]} S$, with $\omega$ in rad/s and the band edge $\Omega = 1/\DEV_0$ set by the finest channel ($[\Omega_{\min},\Omega_{\max}]$ in \cref{scale_distance} is in Hz); the quality factor is
\vspace{-0.1em}
\begin{equation} \label{eq:constant_quality}
    Q = {\omega_{\rm peak}}/{\mathrm{BW}} = {\omega_{\rm peak}}/{(\omega_+ - \omega_-)},
    \vspace{-0.1em}
\end{equation}
with $\omega_\pm$ denoting the $-$3 dB band edges.
The geometric grid \eqref{eq:scale_level_spread} makes every channel a log-shift of the first, $\omega(k) = c^{-(k-1)}\omega(1)$ applied to $\omega_{\rm peak}$ and $\omega_\pm$ alike, so ${\rm BW}(k) = \omega_{\rm peak}(k)/Q(c)$ and the tiling is constant-$Q$, the $K$ channels spanning a peak-frequency ratio $\omega_{\rm peak}(1)/\omega_{\rm peak}(K) = c^{K-1}$.
\Cref{tab:spectral} collects the results.

\subsection{Difference-of-Gaussians} \label{dog}
With $\widehat h_{\rm Gauss}(\omega; \DEV) = e^{-\DEV^2\omega^2/2}$ and $a_k = |\widehat{h}_{\rm Gauss}(\omega;\DEV_k)|$, the energy capture telescopes to
\vspace{-0.1em}
\begin{equation} \label{eq:dog_deficit}
    S_{\rm DoG}(\omega;\DEV_K, c) = a_0^2 - 2\sum_{k=1}^K a_k (a_{k-1} - a_k) \leqslant 1,
    \vspace{-0.1em}
\end{equation}
attaining $B = 1$ at $\omega = 0$.
Each deficit term $a_k (a_{k-1} - a_k)$ vanishes at $\omega = 0$ and as $\omega \to \infty$, so $S_{\rm DoG}$ ripples as $c$ grows and the lower bound $A = \inf_{\omega \in [0, \Omega]} S_{\rm DoG} > 0$ has no closed form in general.
On $[0, \Omega]$, $A = S_{\rm DoG}(\Omega)$ from \eqref{eq:dog_deficit}, giving $B/A = 8.00$ at $c = \sqrt{2}$, independently of $K$. The DoG forms an overcomplete, non-tight frame on bounded bands, mirroring the frame analysis of decimated Laplacian pyramids \cite{do2003framing}.
Its peak is closed form, from $\frac{\mathrm d}{\mathrm d\omega}|\widehat{\psi}_{\rm DoG}|^2 = 0$ with $\sigma_k = c\,\sigma_{k-1}$,
\vspace{-0.1em}
\begin{equation} 
    \label{eq:dog_peak}
    \omega_{\rm peak}^{\rm DoG} = {2}{\sigma_{k-1}^{-1}} \, \sqrt{{\ln c}/{c^2 - 1}},
    \vspace{-0.1em}
\end{equation}
while the $-$3 dB edges solve a difference of two Gaussians and are found numerically.
The channel is a difference of Gaussians, so it decays as $e^{-\DEV_{k-1}^2\omega^2/2}$.

\vspace{-0.3em}
\subsection{Difference-of-time-causal-limit-kernels} 
\label{dot}
\vspace{-0.2em}

Let $\Psi$ be the time-causal limit kernel, $h_\Psi$ \eqref{eq:time-causal_limit_kernel_convolution}.
It is an admissible mother wavelet \cite[Sec.~2.5.7]{lindeberg2025time}, with Laplace transform \cite[(37)]{lindeberg2025time}
\vspace{-0.3em}
\begin{equation} \label{eq:time-causal_limit_kernel_laplace}
    \widehat{\Psi}(s; \DEV_k, c) = \prod_{j=1}^\infty \frac{1}{1 + \TIMECONSTANT_j s}, \qquad \TIMECONSTANT_j = c^{-j} \sqrt{c^2 - 1} \, \DEV_k.
    \vspace{-0.4em}
\end{equation}
Using $\sigma_k=c \, \sigma_{k-1}$ \eqref{eq:scale_level_spread}, shifts the product index by one scale level to reduce the DoT difference to one factor in the Fourier domain ($s = i\omega$)
\begin{equation} \label{eq:dot_simplified_fourier}
    |\widehat{\psi}_{\rm DoT}(\omega; \DEV_k, c)|^2
     = (c^2-1)\, {\DEV_{k-1}^2} \, \omega^2\ \left|\widehat{\Psi}(\omega; \DEV_k, c)\right|^2 .
\end{equation}
The energy capture, whose lowpass term keeps $A$ from vanishing at $\omega = 0$, is $S_{\rm DoT} = |\widehat{\Psi}(\omega;\sigma_K,c)|^2 + \sum_{k=1}^K |\widehat{\psi}_{\rm DoT}(\omega;\DEV_k,c)|^2$.
Abbreviating $\widehat{\Psi}_k \equiv \widehat{\Psi}(\omega; \DEV_k, c)$ and $\beta_k = \sqrt{c^2-1} \, {\sigma_{k-1}} \, \omega$, each factor of \eqref{eq:time-causal_limit_kernel_laplace} gives $|\widehat{\Psi}_{k}|^2 = |\widehat{\Psi}_0|^2/\prod_{j=1}^{k}(1 + \beta_j^2)$, so the band sum telescopes and the energy capture reduces to the finest scale alone,
\begin{equation} \label{eq:psi_dot_k-th}
    S_{\rm DoT} = |\widehat{\Psi}_K|^2 + |\widehat{\Psi}_0|^2 - |\widehat{\Psi}_K|^2 = |\widehat{\Psi}_0|^2 .
\end{equation}
Each factor of $|\widehat{\Psi}_0|^2$ is 1 at $\omega = 0$ and decays monotonically, so $B = 1$ and, on $\omega \in [0, 1/\DEV_0)$, $A = \prod_{j=1}^\infty (1 + c^{-2j}(c^2-1))^{-1}$ depends on $c$ alone: both bounds are independent of $K$, giving $A = 0.419$, $B/A = 2.38$ at $c = \sqrt{2}$, an overcomplete non-tight frame.
The peak has no simple closed-form solution.
Setting $\frac{\mathrm d}{\mathrm d\omega}\log|\widehat{\psi}_{\rm DoT}|^2 = 0$ in \eqref{eq:dot_simplified_fourier} and substituting $x = \omega^2\mu_1^2$ gives the condition
\begin{equation} \label{eq:dot_log_sum}
    \sum_{j=1}^\infty \frac{x \, c^{-2(j-1)}}{1 + x \, c^{-2(j-1)}} = 1 ,
\end{equation}
whose unique solution $x^*(c)$, bounded by $(c^2-1)/c^2 \leqslant x^*(c) \leqslant c^2-1$, pins the peak to
\begin{equation} \label{eq:dot_peak_sandwich}
    {\DEV_k^{-1}} \leqslant \omega_{\rm peak}^{\rm DoT} \leqslant {c}{\DEV_k^{-1}} .
\end{equation}
The $N$-fold cascade decays as $O(\omega^{-N})$, i.e. $6N$~dB/octave.

\subsection{Difference-of-truncated-exponentials} \label{doe}
Each DoE channel uses a single truncated exponential \eqref{eq:truncated_exponential} per scale (the impulse response of a first-order integrator \cite[(1.11)]{gerstner2014neuronal}, i.e. one recursive filter per channel) with the Laplace transform $\mathcal{L}\{h_{\rm exp}\}(s) = (1 + \TIMECONSTANT s)^{-1}$, so the channel magnitude, with $\nu = \TIMECONSTANT_k^2\omega^2$, is
\begin{equation} \label{eq:doe_squared_magnitude}
    \begin{aligned}
        |\widehat{\psi}_{\rm DoE}|^2
        &= \frac{c-1}{c+1}\left(\frac{1}{1 + c^{-2}\nu} - \frac{1}{1 + \nu}\right) \\[-0.2em]
        &= \frac{\nu \, (1/c - 1)^2}{1 + \nu  (1 \!+\! c^{-2}) + \nu^2c^{-2}}.
    \end{aligned}
\end{equation}
Cancelling terms in the channel sum gives the energy capture with the coarsest lowpass term
\begin{equation} \label{eq:doe_fourier_with_lowpass}
    S_\mathrm{DoE}
    = \frac{1}{c+1}\left(\frac{2}{1 + \TIMECONSTANT^2_K\omega^2} + \frac{c-1}{1 + c^{-2}\TIMECONSTANT_1^2\omega^2}\right) ,
\end{equation}
monotonically decreasing from $B = S_\mathrm{DoE}(0) = 1$.
On $\omega\in[0,1/\TIMECONSTANT_0]$, the infimum
\begin{equation} \label{eq:doe_lower_bound}
    A = \frac{2}{(c+1)(1 + c^{2K})} + \frac{c-1}{2(c+1)} \;\xrightarrow[K \to \infty]{}\; \frac{c-1}{2(c+1)}
\end{equation}
gives $A = 0.086$, $B/A = 11.63$ at $c = \sqrt{2}$ and $K = 12$, a non-tight frame which, unlike the DoT, does not converge exactly --- the lowpass term vanishes only as $K\to\infty$, where $B/A \to 11.66$.
Differentiating \eqref{eq:doe_squared_magnitude} gives the exact peak at $\nu = c$,
\begin{equation} \label{eq:doe_peak}
    \omega_{\rm peak}^{\rm DoE} = \frac{\sqrt{c}}{\TIMECONSTANT_k} = \frac{1}{\sqrt{\TIMECONSTANT_{k-1} \, \TIMECONSTANT_k}},
\end{equation}
the geometric mean of adjacent time constants.
Setting $|\widehat\psi_{\rm DoE}|^2 = \tfrac12|\widehat\psi_{\rm DoE}|^2_{\max}$ with the peak value $(c-1)^2/(c+1)^2$ from \eqref{eq:doe_squared_magnitude} and clearing the denominator, the $-$3 dB condition reduces to the biquadratic $\nu^2 - (c^2+4c+1)\nu + c^2 = 0$, with roots
\begin{equation} \label{eq:doe_3db_roots}
    \nu_\pm = \tfrac{1}{2}\left[(c^2 + 4c + 1) \pm (c+1)\sqrt{c^2 + 6c + 1}\right],
\end{equation}
so the fractional bandwidth $(\sqrt{\nu_+} - \sqrt{\nu_-})/\sqrt{c}$ depends only on $c$.
A single channel decays as $O(\omega^{-1})$, i.e. $6$~dB/octave.

\begin{table}[t]
    \vspace{-0.6em}
    \centering
    \caption{Spectral characterization at $c = \sqrt{2}$, $K = 12$ channels for the DoE bound, and the $N \to \infty$ limit kernel for the DoT, save the decay row. Conditioning $B/A$ on the common band spanned by the finest channel. $-$3 dB edges of that channel, scaled by the finest scale ($\DEV_0$, resp. $\TIMECONSTANT_0$). $\ast$ denotes numerically evaluated quantities.}
    \label{tab:spectral} % TODO: quote equations with numbers (below (18))
    \begin{tabular}{@{}lccc@{}}
        \toprule
                       & DoG & DoT & DoE \\
        \midrule
        $B/A$          & $8.00^{\ast}$ \eqref{eq:dog_deficit} & $2.38$ (below \ref{eq:psi_dot_k-th}) & $11.63$ \eqref{eq:doe_fourier_with_lowpass} \\
        Peak $\omega_{\rm peak}$ & Closed \eqref{eq:dog_peak} & Bracket \eqref{eq:dot_peak_sandwich} & Closed \eqref{eq:doe_peak} \\
        $-$3 dB $\omega_\pm$ & $0.72, 1.72^{\ast}$ & $0.38, 1.59^{\ast}$ & $0.34, 2.05$ \eqref{eq:doe_3db_roots} \\
        $Q$ \eqref{eq:constant_quality} & $1.18^{\ast}$ & $0.70^{\ast}$ & $0.49$ \\
        Decay          & $e^{-\DEV^2\omega^2/2}$ & $O(\omega^{-N})$ & $O(\omega^{-1})$ \\
        \bottomrule
    \end{tabular}
    \vspace{-0.6em}
\end{table}

\section{Frame geometries} \label{frame_geometries}

\begin{figure*}[t]
    \centering
    \vspace{-1mm}
    \includegraphics[width=0.95\linewidth]{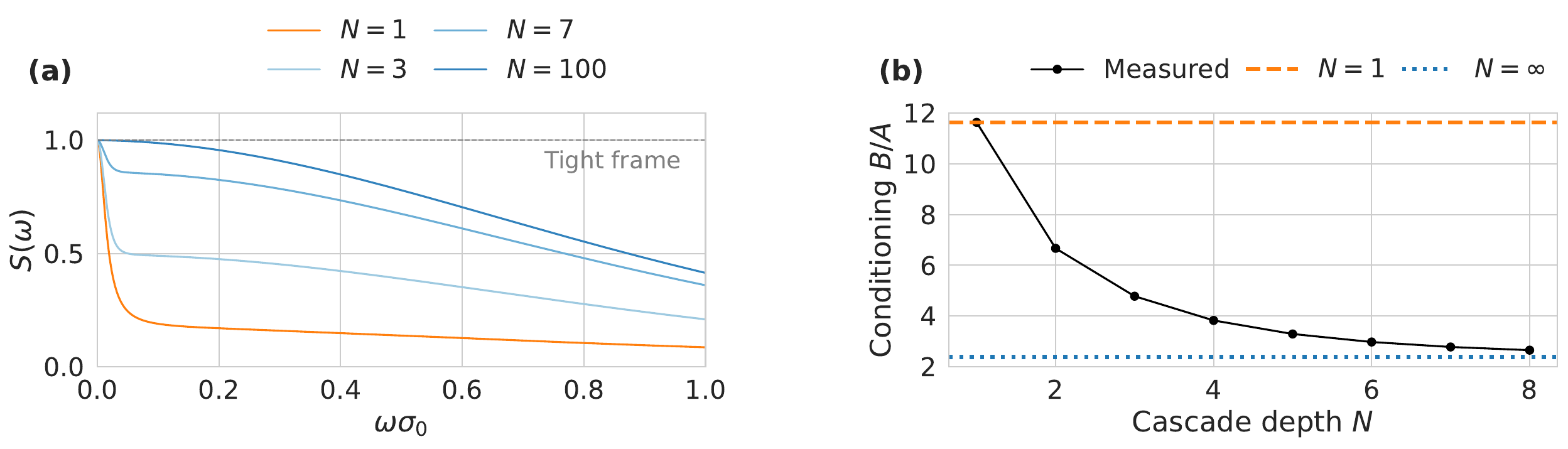}
    % \vspace{-2mm}
    \caption{Implementations of the time-causal bandpass wavelets ($c = \sqrt{2}$, $K = 12$) at varying $N$ (DoE: $N=1$, DoT: $N=\infty$).
    (a) energy capture $S(\omega)$ \eqref{eq:energy_capture_def} for the whole filterbank per cascade depth $N$ relative to the finest scale $\sigma_0$ (a tight frame has $S(\omega) = 1$).
    (b) conditioning versus $N$ at $\sigma_0/\Delta t = 64$, interpolating from DoE \eqref{eq:doe_fourier_with_lowpass} to the DoT \eqref{eq:psi_dot_k-th} limit.
    % (c) relative mismatch between the realized conditioning $B/A$ \eqref{eq:frame_bounds} and its continuous reference (the DoE closed form at $N=1$ \eqref{eq:doe_fourier_with_lowpass} and the $N \to \infty$ limit kernel \eqref{eq:time-causal_limit_kernel_convolution} otherwise \eqref{eq:psi_dot_k-th}) at varying sampling steps.
    }
    \label{fig:exp_discretization}
    \vspace{-1mm}
\end{figure*}

The frame bounds $A$ and $B$ established above for the DoG, DoT, and DoE characterize each wavelet channel individually and we now turn to study interacting channels.
\vspace{-0.4em}
\subsection{Channel overlap and Gram matrix structure} \label{gram_matrix}
The quality of a wavelet frame depends not only on the energy capture function $S(\omega)$ from above, but also on the overlap between the channels.
The Gram matrix $\boldsymbol{G} \in \mathbb{R}^{K \times K}$ has the entries
\begin{equation}
    G_{jk} = \langle \psi_j, \psi_k\rangle = \int_{-\infty}^{\infty} \widehat{\psi}_j(\omega) \, \overline{\widehat{\psi}_k(\omega)} \, {\rm d}\omega,
\end{equation}
i.e., the matrix of the Gram operator $\Lambda\Lambda^*$ restricted to the $K$ channels, which shares its nonzero spectrum with $\Lambda^*\Lambda$.
Since the wavelets are real-valued, the integrand is conjugate symmetric and $G_{jk} = 2\int_0^{\infty} {\rm Re}[ \widehat{\psi}_j(\omega) \, \overline{\widehat{\psi}_k(\omega)}] \, {\rm d}\omega$, characterizing the geometry of the frame.
The normalized Gram matrix
\begin{equation} \label{eq:gram_normalized}
    G_{{\rm norm}, jk} = G_{jk} / \sqrt{G_{jj} \, G_{kk}},
\end{equation}
where $G_{jj} = \langle \psi_j, \psi_j \rangle = \|\psi_j\|^2$, is the identity precisely when the channels are orthogonal.
Tightness ($\Lambda^*\Lambda = A\,\boldsymbol{I}$) only constrains the aggregate overlap $S(\omega)$, so off-diagonals might appear even for tight frames.

The off-diagonal entries of $G_{\rm norm}$ reveal the spectral overlap between the channels, and the condition number ${\rm cond}(G) = B/A$ measures how far the frame is from being tight.
A large ${\rm cond}(G)$ means that the reconstruction is sensitive to noise and quantization, while the rapid off-diagonal decay in $G_{\rm norm}$ (a small ${\rm cond}(G)$) means that each channel captures information that is approximately independent of distant channels.

\vspace{-0.3em}
\subsection{Choosing the distance between scales} \label{scale_distance}
In this context, the scale distance parameter $c$ governs the trade-off between dense scale sampling and improved frequency coverage vs.\ increased overlap and worse ${\rm cond}(G)$.
The DoE channels overlap broadly with a relatively slow spectral decay compared to the DoT (\cref{fig:bandpass}).
Increasing the cascade order $N$ sharpens the bandpass response (\cref{fig:exp_discretization}, a), and reduces the overlap between adjacent channels and hence the off-diagonal mass of $G_{\rm norm}$.

\vspace{-0.4em}
\subsubsection{Closed form for \textit{c} from the frequency range}
\vspace{-0.2em}
With $K$ channels covering a target frequency range $[\Omega_{\min}, \Omega_{\max}]$ in $\mathrm{Hz}$, the geometric spacing of the variances $\tau_k = \tau_{\max} \, c^{-2(K-1-k)}$ together with $f \approx 1/(2\pi \sqrt{\tau})$ fixes $c$ uniquely:
\begin{align}
    \tau_{\max} &= {(2\pi \, \Omega_{\min})^{-2}}, \qquad
    \tau_{\min} = {(2\pi \, \Omega_{\max})^{-2}},
\\
\label{eq:c_from_freq_range}
    c &= \left({\tau_{\max}}/{\tau_{\min}}\right)^{\!1/(2(K-1))}
    = \left({\Omega_{\max}}/{\Omega_{\min}}\right)^{\!1/(K-1)}.
\end{align}
Equation~\eqref{eq:c_from_freq_range} exposes the trade-off: holding $[\Omega_{\min}, \Omega_{\max}]$ fixed and increasing $K$ drives $c \to 1$, packing channels more densely but increasing pairwise overlap (off-diagonal entries of $G_{\rm norm}$) and worsening the conditioning of $\Lambda^*\Lambda$.
\vspace{-0.4em}
\subsubsection{Stability constraints on \textit{c}}
\vspace{-0.2em}
Two practical lower bounds on $c$ enter the implementation based on numerical leaky-integrator stability and cascade order.
The discrete first-order integrator with time constant $\TIMECONSTANT$ has decay $\alpha = \exp(-\Delta t / \TIMECONSTANT)$.
Requiring $\alpha \geq \alpha_{\rm floor}$ for some floor (we use $\alpha_{\rm floor} = 0.01$) imposes $\TIMECONSTANT \geq \Delta t / (-\ln \alpha_{\rm floor})$ at every cascade stage, which through the discrete recursion $\TIMECONSTANT = (-1+\sqrt{1+4\Delta\tau})/2$ \cite[(58)]{lindeberg2016timecausal} translates into a minimum $\Delta\tau$ at the finest stage.
For DoT, this constraint propagates to a maximum cascade order $N^\star$ given $K$ and $c$: when the $c$ chosen by \eqref{eq:c_from_freq_range} would push the finest stage below $\alpha_{\rm floor}$, we reduce $N$ until each stage is stable, accepting a less smooth limit-kernel approximation as the price of frequency coverage.
A second bound enforces channel distinctness: as $c \to 1$, the bandpass amplitude $\lVert\psi_k\rVert$ collapses (adjacent low-pass kernels nearly cancel).
We therefore floor $c \geq 1.05$, ensuring that channels stay meaningfully distinct.
These floors set a minimum $c_{\min}$ and \eqref{eq:c_from_freq_range} implies a maximum number of useful channels for a given frequency range
\vspace{-0.1em}
\begin{equation} \label{eq:k_max}
    K_{\max} = 1 + \left\lfloor {\log(\Omega_{\max}/\Omega_{\min})}/{\log c_{\min}} \right\rfloor.
    \vspace{-0.1em}
\end{equation}
% The stability bounds and spectral characteristics assume continuous time and, for the DoT, an infinite cascade ($N \to \infty$).
% We compare the conditioning $B/A$ between the discrete recursive-filter implementation and the continuous closed forms over varying sampling steps $\Delta t$ (\cref{fig:exp_discretization}, b), measured as relative distance.
% For DoE the reference is exact, so the deviation stems exclusively from discretization error and decays as $\Delta t^2$.
% For DoT, the reference is the $N \to \infty$ limit kernel \eqref{eq:time-causal_limit_kernel_convolution}, so the deviation also carries a finite-$N$ truncation term that is constant in $\Delta t$.
Sweeping $N$ (\cref{fig:exp_discretization}, c) shows the realized conditioning interpolating monotonically from the DoE closed form at $N = 1$ toward the DoT limit.

\vspace{-0.4em}
\section{Conclusion}
We introduced the Difference-of-truncated-Exponentials (DoE) and characterized it, the DoT, and the non-causal DoG baseline in terms of frame bounds and conditioning, peak frequencies, constant-$Q$ bandwidth tiling, and spectral decay.
The DoT is the best-conditioned ($B/A=2.38$ at $c=\sqrt{2}$) but requires $N$ cascaded filters per channel, whereas the DoE trades conditioning ($B/A=11.63$) for a single integrator per channel.

The spectral characterization in \cref{spectral} and the frame geometries in \cref{frame_geometries} provide rules for the scale ratio $c$ from a target band and the channel count $K_{\max}$ \eqref{eq:k_max} with which the causal wavelets can be implemented.

\vfill\pagebreak

\bibliographystyle{IEEEbib}
\bibliography{RefSNN}

\end{document}